\documentclass[aps,prb,twocolumn]{revtex4}
\usepackage{graphicx,subfigure}
\usepackage{bm}
\usepackage{amsfonts}
\usepackage{dsfont}
\usepackage{color}
\usepackage{enumerate}

\begin{document}

\title{Topological Phases of One-Dimensional Fermions: An Entanglement Point
of View}
\author{Ari M. Turner}
\affiliation{Department of Physics, University of California, Berkeley CA 94720, USA}
\author{Frank Pollmann}
\affiliation{Department of Physics, University of California, Berkeley CA 94720, USA}
\affiliation{Institute of Physics, Academia Sinica, Taipei 11529, Taiwan}
\author{Erez Berg}
\affiliation{Department of Physics, Harvard University, Cambridge, MA 02138, USA}
\date{\today}

\begin{abstract}
The effect of interactions on topological insulators and superconductors
remains, to a large extent, an open problem. Here, we describe a framework for classifying phases of one-dimensional interacting
fermions, focusing on spinless fermions with time-reversal symmetry and
particle number parity conservation, using concepts of entanglement. In
agreement with an example presented by Fidkowski \emph{et. al.} (Ref. [%
\onlinecite{Fidkowski-2010}]), we find that in the presence of interactions
there are only eight distinct phases, which obey a $\mathbb{Z}_8$ group
structure. This is in contrast to the $\mathbb{Z}$ classification in the
non-interacting case. Each of these eight phases is characterized by a
unique set of bulk invariants, related to the transformation laws of its
entanglement (Schmidt) eigenstates under symmetry operations, and has a
characteristic degeneracy of its entanglement levels. If translational
symmetry is present, the number of distinct phases increases to 16.
\end{abstract}

\maketitle

\section{Introduction}

Topological phases of matter are not characterized by a broken symmetry, but
rather by an underlying topological structure that distinguishes them from
other, topologically trivial phases. Such phases have attracted a great deal
of attention recently, especially since the theoretical prediction \cite%
{Kane2005, Bernevig2006,Fu2007,MooreBalents2007} and subsequent experimental
observation \cite{Konig2007,Hasan2008} of both two- and three-dimensional
realizations of time-reversal invariant topological insulators. These (as
well as their predecessor, the integer quantum Hall effect) can be thought
of as band insulators characterized by the topological structure of their
Bloch bands. Similarly, topological superconductors\cite%
{Volovik-1988,Schnyder-2008,Kitaev-2009} are characterized by the
topological nature of their fermionic quasi-particle spectrum. All these
systems can be understood from a non-interacting point of view. A complete
classification of all topological phases of non-interacting fermions, given
their symmetries, has been given in Refs.
\onlinecite {Schnyder-2008,
Qi-2008, Kitaev-2009}.

In the presence of electron-electron interactions, the Hamiltonian cannot be
reduced to a single particle matrix. Therefore, strictly speaking, the above
classification scheme of topological phases cannot be used. Nevertheless, in
some classes of topological insulators, the topological order has been
argued to be robust even in the presence of interactions, by generalizing
the corresponding topological invariant to the many-body case\cite%
{NiuThouless,Avron1985,Qi-2008}. In other classes, however, the situation in
the interacting case remains unclear.

In a recent breakthrough, Fidkowski and Kitaev studied a one-dimensional
model of spinless superconductors with time reversal symmetry.\cite%
{Fidkowski-2010} They found that in the presence of interactions, the
free-fermion classification breaks down from $\mathbb{Z}$ to $\mathbb{Z}_{8}
$, \emph{i.e.} there are only eight distinct phases that survive in the
presence of interactions (as opposed to an infinite number without
interactions). To the best of our knowledge, this is the first case where
the non-interacting picture in a class of topological phases is found to be
radically modified by interactions. Ref. \onlinecite{Fidkowski-2010}
constructs an explicit path in Hamiltonian space through which phases with
different $\mathbb{Z}$ numbers $\mathrm{mod}(8)$ can be connected, and also
discusses the stability of the edge states. However, a more general
understanding of the classification of distinct phases in the presence of
interactions (in particular, in terms of \emph{bulk} properties of the
ground state wavefunction) is left open.

In this paper, we develop a framework for classifying phases of interacting
fermions in one dimension based on bipartite entanglement of the ground states wave function.
The fact that entanglement is a useful quantity to probe topological properties of wave functions has been shown in several recent publications, see for example Refs.~\onlinecite{Kitaev-2006,Levin-2006,Li-2008,Nussinov-2009}.
Our technique is based on a method which was introduced in
Ref. \onlinecite{Pollmann-2010} for classifying phases in spin systems
spins. This method has also been developed
more fully and shown to give a \emph{complete} 1D classification
by Ref. \onlinecite{Chen-2010} (at least when translational symmetry
is not required). Here we generalize the method to fermionic systems.   We find
that the eight phases found in Ref. \onlinecite{Fidkowski-2010} are indeed
topologically distinct, and characterize them in terms of a set of
invariants. These phases cannot be continuously connected by any kind of
interaction as long as time-reversal symmetry and fermion parity
conservation are preserved.

The basic idea is to examine the behavior of the entanglement (Schmidt)
eigenstates of a segment in the bulk of the system under the symmetry group
of the system. Topologically nontrivial phases can be recognized by the
presence of ``fractionalized'' modes in the entanglement spectrum, which
transform differently under the symmetry group from the constituent
microscopic degrees of freedom of the system (analogous to the half-integer
spins at the ends of the spin one Heisenberg chain). The character of the
entanglement spectrum cannot change without a bulk phase transition, at
which the nature of the ground state changes abruptly or the correlation
length diverges.

The behavior of the entanglement modes reflects the character of the \emph{%
physical} topologically-protected modes at the boundary of the system.
However, unlike the edge modes, the entanglement spectrum represent a truly
\emph{bulk} property of the ground state wavefunction, and as such, it is
not sensitive to symmetry-breaking perturbations at the surface.

We start in Sec. \ref{nr1} by introducing fermionic Hamiltonians with
pairing terms, through the example of a single Majorana chain model. The
general framework to classify topological phases based on symmetry
properties of the entanglement eigenstates is presented in Section~\ref%
{sec:sym}. We apply it to fermionic systems with time-reversal invariance
and fermion number parity conservation, and derive the invariants
characterizing the eight distinct phases and the degeneracies in their
entanglement spectrum. These phases are shown in the next section to have a $%
\mathbb{Z}_{8}$ group structure, defined through the rules for combining
phases with different invariants. In Sec.~\ref{sec:Addition}, we demonstrate
how to construct each phase by combining single chains. In Sec. ~\ref%
{sec:trans}, we discuss the additional phases which arise if translational
symmetry is imposed. The results are summarized and discussed in Sec.~\ref%
{sec:sum}.

\section{Fermionic models with pairing terms\label{sec:mod}}

\label{nr1}We will investigate time-reversal invariant one-dimensional
systems of spinless fermions, in which the particle number is conserved
modulo two. (The classification of topological phases is most interesting in
this case.) Such a situation can be realized in a system in contact with a
superconductor. As a simple example, consider the following
Hamiltonian\cite{Kitaev-2001}:
\begin{eqnarray}
H_{0} &=&-\frac{t}{2}\sum_{j}\left( c_{j}^{\dagger }c_{j+1}^{\dagger
}+c_{j}^{\dagger }c_{j+1}^{\vphantom{\dagger}}+\mbox{H.c.}\right)
+u\sum_{j}c_{j}^{\dagger }c_{j}^{\vphantom{\dagger}},  \nonumber
\label{singlechain} \\
&&
\end{eqnarray}%
with $t,u\geq 0$. The operators $c_{j}^{\dagger }$ ($c_{j}^{%
\vphantom{\dagger}}$) create (annihilate) a spinless fermion on site $j$.
The first term comprises hopping of fermions as well as the creation and
annihilation of pairs of fermions while the second term acts as a chemical
potential. The fermion parity operator
\[
Q=e^{i\pi \sum_{j}n_{j}}\label{Q}
\]%
with $n_{j}=c_{j}^{\dagger }c_{j}$ commutes with $H_{0}$ as the total number
of fermions $N_{\text{total}}$ modulo two is conserved. Furthermore, the
Hamiltonian is time-reversal symmetric. (For spinless fermions, time
reversal is represented by complex conjugation.)

Let us begin by considering only the conservation of the fermion number
parity. This symmetry of $H_{0}$ allows us to distinguish two different
phases. The system undergoes phase transitions at $t=u$, but \emph{no} local
order parameter can be used to distinguish the two phases on either side of
the transition. However, they can be distinguished by their edge states. In
the phase $u>t$, the ground state for an open chain is unique while it is
two-fold degenerate for $t>u$.\cite{Kitaev-2001} If $u=0$, one can check
that these states are given by the equal weighted superposition of all
configurations with fixed fermion parity (i.e., an even number or an odd
number of particles). The ground state degeneracy in this case can be
understood in term of degrees of freedom at the two ends. The two ground
states cannot be distinguished by $\emph{any}$ local observable in the bulk,
because in any finite region of either state, the parity can be either even
or odd. However, the two states \emph{are} distinguishable when the opposite
ends are compared to one another: $c_{N}^{\dagger }c_{1}+c_{N}^{\dagger
}c_{1}^{\dagger }+\mbox{H.c.}$ has a different eigenvalue for the two
states. Furthermore, we can transform the two ground states into each other
by acting with either $c_{1}+c_{1}^{\dagger }$ or $c_{N}-c_{N}^{\dagger }$
on the two ends of the chain. In other words, there is a single fermionic
state that is split between the ends of the chain, and the observable
described above measures its occupation number. The two states are
degenerate because the only distinction between them is long-range, while
energy measures only local correlations. In the phase $t<u$, however, there
is no such degeneracy. The picture provided remains true even if we include
interactions as the arguments can be stated in a way that only requires the
Hamiltonian to conserve the fermion parity.

The edge properties have a simple explanation in a different representation
defined by the following transformation:
\begin{eqnarray}
a_{j} &=&c_{j}+c_{j}^{\dagger } \\
b_{j} &=&-i(c_{j}-c_{j}^{\dagger }).
\end{eqnarray}%
$a_{j}$ and $b_{j}$ are Majorana operators; they obey the relations $%
\{a_{i},a_{j}\}=\{b_{i},b_{j}\}=\delta _{ij}$, $\{a_{i},b_{j}\}=0$, $a_{i}^{%
\vphantom{\dagger}}=a_{i}^{\dagger }$ and $b_{i}^{\vphantom{\dagger}%
}=b_{i}^{\dagger }$. The fermion parity, $1-2n_{j}$ of a site, is given by $%
ib_{j}a_{j}$. Using these operators, $H_{0}$ can be written (up to a
constant) as
\begin{equation}
H_{0}=\frac{i}{2}\left( t\sum_{j}b_{j}a_{j+1}+u\sum_{j}a_{j}b_{j}\right) .
\end{equation}%
Observe that each unit cell now contains two operators. In the case where $%
t=0,u=1$, the ground state is described by $ia_{j}b_{j}=-1$, i.e., each site
is vacant. In terms of these variables, the phase $t=1$, $u=0$ is also
simple: $ib_{j}a_{j+1}=-1$ when $t$ is large. (This can be regarded as the
parity of a fermion shared between sites $j$ and $j+1$.) This requirement
does not completely determine the ground state wave function in an open
chain, though, because it leaves $a_{1}$ and $b_{N}$ free. There are
therefore two degenerate states characterized by the occupation of the
fermion shared between the ends, $ib_{N}a_{1}$.

The presence of time-reversal symmetry leads to additional
distinctions between phases. Quadratic, time reversal invariant
fermionic Hamiltonians with conservation of the
fermion number $\mathrm{mod}\left( 2\right) $ have been shown\cite%
{Schnyder-2008,Kitaev-2009} to support phases classified by an integer $n\in
\mathbb{Z}$ (class BDI, according to Ref. \onlinecite{Schnyder-2008}). Each
phase is characterized by having $n$ gapless Majorana modes at each edge,
and the different phases cannot be smoothly connected to each other without
closing the bulk gap. It was later found\cite{Fidkowski-2010} that in the
presence of interactions this classification breaks down to $\mathbb{Z}_{8}$%
. Now we will begin the main discussion, whose goal
is to show how the eight distinct phases
in the general (interacting) case can be understood and classified according
to properties of their entanglement eigenstates under the symmetries of the
system, namely time-reversal and fermion number parity conservation.

\section{Classifying phases by symmetry properties of the entanglement
eigenstates\label{sec:sym}}

\begin{figure}[tbp]
\begin{center}
\includegraphics[width=85mm]{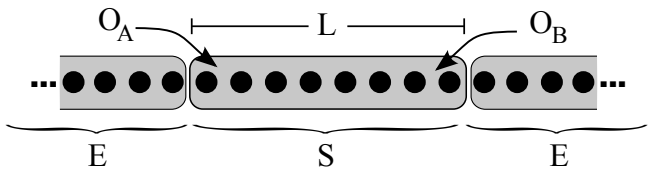}
\end{center}
\caption{Illustration of a bipartition of a 1D chain into a segment (S) of
length L and a surrounding environment (E). The operators $\mathcal{O}_A$
and $\mathcal{O}_B$ act on the edges of the segment.}
\label{fig:block}
\end{figure}

In the preceding section, we discussed physical edge properties to
characterize different phases. Below, we present an alternative method of
classifying the phases, which \emph{involves only the bulk}. This is
achieved by examining the \textquotedblleft entanglement
spectrum\textquotedblright \cite{Li-2008, Thomale-2009,
Fidkowski-2010a,Pollmann-2010, Laeuchli-2010} of a finite (but arbitrarily
large) segment\cite{Levin-2006,Kitaev-2006}, embedded in an infinite system.
The transformation law of the entanglement (or Schmidt) states under the
symmetry group of the system can be used to distinguish between different
phases, as we describe below.

We consider a bipartition of a 1D chain with periodic boundary conditions
into a segment ($S$) of length $L$ and a surrounding environment ($E$) of
length $N\gg L$ as shown in Figure \ref{fig:block}. For the segment $S$, the
reduced density matrix of the ground state wave function $|\psi \rangle $ is
given by
\begin{equation}
\rho _{S}=\mbox{tr}_{E}\left( |\psi \rangle \langle \psi |\right) .
\end{equation}%
It is convenient to define an \textquotedblleft entanglement
Hamiltonian\textquotedblright\ $\mathcal{H}_{S}$ such that
\begin{equation}
\rho _{S}=e^{-\mathcal{H}_{S}}.
\end{equation}%

Then, the low \textquotedblleft energy" states of $\mathcal{H}_{S}$ are the
most likely states of the segment $S$, when the entire system is in its
ground state. We call the eigenstates $|\phi _{\gamma }\rangle _{S}$ of $%
\mathcal{H}_{S}$ \textquotedblleft entanglement eigenstates" and the
eigenvalues $E _{\gamma }$ \textquotedblleft entanglement eigenvalues".
(These are the same as the Schmidt states defined by $|\psi \rangle =\sum
e^{-\frac{E _{\gamma }}{2}}|\phi _{\gamma }\rangle _{S}|\chi _{\gamma
}\rangle _{E}$, where $|\chi _{\gamma }\rangle _{E}$ are the corresponding
Schmidt states of the environment.)

Our approach is based on an important observation for the
entanglement Hamiltonian  $\mathcal{H}_{S}$: The low
``entanglement energy'' excitations of $\mathcal{H}_{S}$ in a
$d$-dimensional system may be well described by a $d-1$
dimensional effective Hamiltonian (see also Ref.
\onlinecite{OsborneHolog} for an interesting discussion of this
concept). We consider the entanglement spectrum of a sufficiently
large one-dimensional segment $S$, and focus on the ``low-lying"
entanglement states with $E_{\gamma}<E_{\text{cut}}$  (where
$E_{\text{cut}}$ is an arbitrary constant).

We now make a crucial observation: in a gapped
system with a finite correlation length $\xi $,\cite{Hastings-2006} these
states can only be distinguished from each other by their behavior within a
certain distance of the ends of the segment $S$. This is justified by the
following argument: suppose that we measure a correlation function $C(\ell
)=\langle \mathcal{O} _{E}\mathcal{O}_{S}\rangle _{\psi }$, where $\mathcal{O%
}_{E}$ acts on sites in the environment $E$ and $\mathcal{O}_S$ acts on
sites in the segment (far away from the edges), respectively. $\ell $ is the
minimal distance between the sites on which $\mathcal{O}_{E}$ and $\mathcal{O%
}_{S}$ act. We expect that as $\ell $ becomes large, $C(\ell )\rightarrow
\langle \mathcal{O}_{E}\rangle _{\psi }\langle \mathcal{O}_{S}\rangle _{\psi
}$. Let us take for $\mathcal{O}_{E}$ an operator that projects onto a
particular Schmidt state of the environment, say $|\chi _{\gamma }\rangle_E $%
. ($\mathcal{O}_S$ can be any operator as long as it acts far from the edges
of $S$.) When applied to the ground state, $\mathcal{O}_E$ projects (through
the entanglement between $E$ and $S$) also onto the corresponding
entanglement eigenstate $|\phi _{\gamma }\rangle _{S}$ of $S$. Thus, we have
$\langle \mathcal{O}_{S}\rangle _{\phi _{\gamma }}\approx \langle \mathcal{O}%
_{S}\rangle _{\psi }$. That is, $\langle \mathcal{O}_{S}\rangle $ in each
eigenstate of $\mathcal{H}_{S}$ is the same as in the ground state; \emph{%
i.e.}, far enough from the edge of $S$, \emph{any eigenstate of $\mathcal{H}%
_{S}$ behaves essentially like the ground state.} Therefore, an operator
acting far from the edge of $S$ cannot distinguish different Schmidt states;
its expectation value must be the same in all of them. (A simple
generalization of the above argument, using an off-diagonal $\mathcal{O}_{E}$%
, shows that $\mathcal{O}_{S}$ cannot connect different low-energy
entanglement eigenstates if it is sufficiently far from the boundary.)

Furthermore, one can show (see App.~\ref{app:operators} for the case
of bosons) that
any linear transformation applied to the subspace of low-entanglement energy
eigenstates of $S$ can be represented by a local operator acting on sites
within a distance $\ell $ from either boundary of $S$, with an accuracy that
improves exponentially with $\ell $. In particular, every \emph{symmetry
operation} of the system is a unitary operation acting on the entanglement
eigenstates,\cite{PerezGarcia-2008} and thus can be represented (within the
low-entanglement basis) as a product of two operators $\mathcal{O}_{A}$, $%
\mathcal{O}_{B}$ (Fig. \ref{fig:block}), acting on sites near the left and
right edges of $S$, respectively. (Note that $\mathcal{O}_{A}$, $\mathcal{O}%
_{B}$ are specific for a particular ground state $|\psi \rangle $.) Thus, $%
\mathcal{O}_{A}$ and $\mathcal{O}_{B}$ form a (projective) representation of
the symmetry group. Classifying the representations formed by $\mathcal{O}%
_{A}$, $\mathcal{O}_{B}$ distinguishes different phases, which cannot be
adiabatically connected unless a phase boundary is crossed. At the phase
boundary, either the character of the ground state changes discontinuously,
or else the correlation length diverges and the two ends are no longer
independent.

Let us demonstrate these principles through the case of an SU$(2)$-symmetric
integer spin chain. If the segment $S$ is sufficiently long, there can be
situations in which entanglement states come in degenerate sets\cite%
{Pollmann-2010} This can be seen from the fact that, according to the
argument above, any SU$(2)$ rotation can be represented accurately (within
the low-entanglement energy subspace) in terms of two generators, $\mathbf{S}%
_{A}$ and $\mathbf{S}_{B}$, which act only within a distance $\ell$ of the
left and right edges of $S$, respectively. Within this subspace, $\mathbf{S}%
_{tot}\sim \mathbf{S}_{A}+\mathbf{S}_{B}$, where $\mathbf{S}_{tot}$ is the
total spin. Both $\mathbf{S}_{A}$, $\mathbf{S}_{B}$ can be
block-diagonalized into irreducible representations of SU$(2)$ with
well-defined angular momenta. Now, since $\mathbf{S}_{tot}$ is an integer
spin, there are two possibilities: either all the blocks in $\mathbf{S}_{A}$
and $\mathbf{S}_{B}$ form integer representations, or they all form
half-integer representations. This distinguishes two phases: a
\textquotedblleft trivial\textquotedblright\ phase in which $\mathbf{S}_{A}$%
, $\mathbf{S}_{B}$ are both integer (\emph{e.g.}, a fully dimerized chain),
and a \textquotedblleft non-trivial\textquotedblright\ phase in which they
are half-integer (such as the Haldane phase of the spin-one chain\cite%
{Haldane-1983}). In the latter phase, all the entanglement energies of $S$
must have a degeneracy of at least four\cite{Pollmann-2010}, due to the even
degeneracy at each end, guaranteed by the presence of the half-integer spin
operators $\mathbf{S}_{A} $ and $\mathbf{S}_{B}$. This is an alternative
explanation of the symmetry--protection of the Haldane phase, discussed in
Refs. \onlinecite{Hirano-2008,Berg-2008,Gu-2009,
Pollmann-2009a,Pollmann-2010}.

\subsection{Fermion parity}

\label{parrrr} We now turn to discuss fermionic systems. Let us consider a
Hamiltonian $H$ that conserves the fermion parity $Q$ as defined in (\ref{Q}%
) with $Q^{2}=\mathds{1}$. We show that one can distinguish two phases, a
\textquotedblleft trivial\textquotedblright\ phase and a \textquotedblleft
non-trivial\textquotedblright\ one. In the \textquotedblleft
non-trivial\textquotedblright\ phase, a segment's entanglement spectrum is
doubly degenerate. The double degeneracy is related to a single fermionic
degree of freedom, which is split between the opposite ends of the segment.

Any eigenstate of $H$, and in particular the ground state $|\psi \rangle $,
is also an eigenstate of $Q$. Hence the resulting reduced density matrix $%
\rho _{S}$ and the entanglement Hamiltonian $\mathcal{H}_{S}$ both commute
with $Q$. The eigenstates $|\phi _{\gamma }\rangle $ of $\mathcal{H}_{S}$
may therefore be classified by their $Q$-eigenvalues ($q^{S}=+1$ if the
fermion number $N_{F}$ is even and $q^{S}=-1$ if $N_{F}$ is odd.)

We now posit that it is possible to find an effective expression
for $Q$ within the low-entanglement energy subspace, of the form $Q\propto
Q^{A}Q^{B}$, where $Q^{A}$ and $Q^{B}$ are local operators which act near
the left and right edges of $S$, respectively. This is analogous to the
example of spin described in the previous section, in which the total
spin can be represented as a sum of operators acting on the left and right
edges. [In the fermionic case, the decoupling of $Q$ is multiplicative,
since $Q$ is itself a unitary symmetry rather than an SU$\left( 2\right) $
generator.]

Now, $Q^{A}$ and $Q^{B}$ can have interesting relationships: $Q^{A}$ and $%
Q^{B}$ may be two fermionic operators (that is, each contains an odd number
of creation or annihilation operators) or they may both be bosonic. Note
that $Q^{A}$ or $Q^{B}$ cannot contain a sum of bosonic and fermionic terms,
becasue $Q\propto Q^{A}Q^{B}$ is bosonic.\footnote{%
Note that it is not possible to make a fermionic operator out of an even
number of fermionic factors. Why then is it possible to make half-integer
spins out of integer spins in a spin chain? The point is that operators
cannot transform in a ``fractional way."
In fact, the spin operators continue to have the ordinary vector symmetry:
they \emph{change} the spin by integer values, for example.
It is the eigen\emph{states} and not the operators that have half-integer
values. their symmetry properties; only states transform strangely.}

If $Q^{A}$, $Q^{B}$ contain an odd number of fermionic operators, $%
Q^{A}Q^{B}=-Q^{B}Q^{A}$. To emphasize this \textquotedblleft statistical
correlation" between the two ends define an angle $\mu =0$ or $\pi $ to
distinguish between the two cases, so that
\begin{equation}
Q^{A}Q^{B}=e^{i\mu }Q^{B}Q^{A}\text{.}
\end{equation}
For $\mu =0$ ($\pi $), $Q=Q^{A}Q^{B}$ ($-iQ^{A}Q^{B}$). The factor of $i$
must be introduced in the latter case for consistency with the
anticommutation of $Q^{A}$ and $Q^{B}$, given that $Q^{2}=\left(
Q^{A}\right) ^{2}=\left( Q^{B}\right) ^{2}=\mathds{1}$.

For example, consider the Hamiltonian in Eq. \ref{singlechain} with $u=0$, $%
t=1$. In this case, it is not difficult to show that there are only two
entanglement eigenstates on the segment $S$ (with a nonzero weight in the
density matrix), the same two states as the ground states of the original
Hamiltonian restricted to this segment. The fermion parity of these states
is given by
\begin{equation}
Q=-iQ^{A}Q^{B}\text{,}
\end{equation}%
where
\begin{eqnarray}
Q^{A} &=&c_{1}+c_{1}^{\dagger }  \nonumber \\
Q^{B} &=&-i(c_{L}-c_{L}^{\dagger }).
\end{eqnarray}%
We see immediately that $Q^{A}$ and $Q^{B}$ anti-commute, and therefore $\mu
=\pi $. Furthermore, $\left[ \mathcal{H}_{S},Q^{A}\right] =\left[ \mathcal{H}%
_{S},Q^{B}\right] =0$.

For any system with $\mu =\pi $, the above commutation relations
imply that all the eigenvalues of $\mathcal{H}_{S}$ come in
degenerate pairs. To see this, note that $Q$ and $\mathcal{H}_{S}$
can be diagonalized simultaneously. Then, if $\mathcal{H}_{S}|\phi
_{\lambda }\rangle =E _{\lambda }|\phi _{\lambda }\rangle $ and
$Q|\phi _{\lambda }\rangle =q_{\lambda }|\phi _{\lambda }\rangle $
(where $q_{\lambda }=\pm 1$), then the state $|\psi _{\lambda
}\rangle =Q^{A}|\phi _{\lambda }\rangle $ is such that
$\mathcal{H}_{S}|\psi _{\lambda }\rangle =E_{\lambda }|\psi
_{\lambda }\rangle $ and $Q|\psi _{\lambda }\rangle =-q_{\lambda
}|\psi _{\lambda }\rangle $, \emph{i.e.} $|\psi _{\lambda }\rangle
$ is an independent
eigenstate with eigenvalue $E_{\lambda }$. Indeed, for the Hamiltonian (\ref%
{singlechain}) with $u=0$, we find a doubly degenerate entanglement level
with $E =\log 2$. (All other entanglement levels in that system have $E
=\infty $.)

Note that, unlike the Haldane chain example in Sec. \ref{sec:sym}, the
entanglement spectrum is two-fold (rather than four-fold) degenerate. This
is a consequence of the fact that the degeneracy is not associated with
either $Q^{A}$ or $Q^{B}$ alone; it is related to the occupation of the
fermionic level formed by combining $Q^{A}+iQ^{B}$, \emph{i.e.}, it is \emph{%
shared} between the two edges.

In a bosonic system, the states of the entanglement eigenstates can be
represented by $|\phi _{\gamma }\rangle =|\alpha \beta \rangle $. Here $%
\alpha ,\beta $ describe the states of the left and right ends of the chain;
that is, they enumerate the eigenvalues of certain low-\textquotedblleft
energy" combinations of observables that are functions of $\ell $ sites at
the respective ends. This factorization is possible for fermionic chains
with $\mu =0$ as well. However, for chains with $\mu =\pi $, the extra $%
q_{S} $ variable describing the two-fold degeneracy cannot be written in
terms of local observables belonging to either end (\emph{i.e.}, the
fermionic degrees of freedom cannot be measured independently at the two
ends of the segment). Therefore, the entanglement states should be labelled
by three variables, $|\alpha \beta q_{S}\rangle $, with the fermion parity $%
q_{S}$ of the entire chain represented explicitly.\footnote{%
In a chain with $\mu=\pi$, operators on both sides of the chain can change $%
q_S$. The distinction between operators acting on the left and those acting
on the right end is that they are proportional to $\sigma _{x},\sigma _{y}$
respectively. Though these act on the same variable $q_S$, they anticommute.}

In a noninteracting system, the entanglement Hamiltonian is also
noninteracting.\cite{Peschel2004,Cheong-2004} It can be represented in terms
of entanglement modes. The only subtlety is that some of these modes may be
Majorana modes, which satisfy $m^{2}=1;\ m^{\dagger }=m$. There are two
topologically distinct phases, depending on whether there are an even or an
odd number of Majorana modes at each end of the segment. The $Q^{A}$ and $%
Q^{B}$ operators defined above can be found explicitly. Given that the left
edge has $N_{f}$ low--energy fermionic entanglement modes, $f_{A,\alpha }$ ($%
\alpha =1,\dots ,N_{f}$), and $N_{m}$ Majorana modes, $m_{A,\beta }$ ($\beta
=1,\dots ,N_{m}$), $Q^{A}$ is given by
\begin{equation}
Q^{A}=\left( \prod_{\alpha }\left( -1\right) ^{f_{A,\alpha}^{\dagger }f^{%
\vphantom{\dagger}}_{A,\alpha }}\right) \left( \prod_{\beta }m_{A,\beta
}\right) \text{,}
\end{equation}%
and similarly for $Q^{B}$. Note that $Q^{A}$ is a bosonic operator if $N_{m}$
is even (corresponding to $\mu =0$) and a Majorana operator if $N_{m}$ is
odd ($\mu =\pi $).

When interactions are included, only the \textquotedblleft total parities" $%
Q^{A}$, $Q^{B}$ are well-defined. The separate modes $f_{\alpha }$, $%
m_{\beta }$ can \textquotedblleft decay" into other combinations of modes,
but their total is closely related to the symmetry $Q$, as we have just
outlined.

\subsection{Time Reversal Symmetry}

We shall now examine the consequences of time-reversal symmetry on the
degeneracies of the entanglement energies. It turns out that the combination
of time-reversal and fermion parity conservation can have non-trivial
effects.

\paragraph*{Bosonic models.}

Let us introduce the approach by reiterating the results for a bosonic chain
in the presence of time reversal symmetry with $[\mathcal{H}_{S},T]=0$. The
eigenstates $|\phi _{\gamma }\rangle $ can be represented by $|\phi _{\gamma
}\rangle =|\alpha \beta \rangle $, where $\alpha ,\beta $ enumerate the
low-\textquotedblleft energy" states associated with the two edges.
Entanglement eigenstates which differ in their $\alpha $ ($\beta $) index
can be connected by a local operator close to the left (right) edge,
respectively. The transformation of the eigenstates of $\mathcal{H}_{S}$
factors into parts referring to the two ends (see App.~\ref{app:operators}).

It can therefore be represented as a product of two unitary transformations $%
U^{A},U^{B}$ acting on the ends of the segment, so that
\begin{equation}
T|\alpha ,\beta\rangle =\sum_{\alpha ^{\prime },\beta ^{\prime }}U_{\alpha
^{\prime }\alpha}^{A}U_{\beta ^{\prime }\beta}^{B}|\alpha ^{\prime }\beta
^{\prime }\rangle  \label{timefactorization}
\end{equation}
and $[U^{A},U^{B}]=0$.
For a discussion of subtleties related to the anti-unitarity of $T$,
see App.~\ref{app:qs}.
Applying $T$ to an eigenstate twice yields
\begin{eqnarray}
T^{2}|\alpha \beta \rangle &=&\sum_{\alpha ^{\prime },\beta ^{\prime
},\alpha ^{\prime \prime },\beta ^{\prime \prime }}U_{\alpha^{\prime\prime}
\alpha ^{\prime }}^{A}U_{\beta^{\prime\prime} \beta ^{\prime
}}^{B}(U^{A})_{\alpha ^{\prime } \alpha}^{\ast }(U^{B})_{\beta ^{\prime
}\beta} ^{\ast }|\alpha ^{\prime \prime }\beta ^{\prime \prime }\rangle .
\nonumber \\
&&  \label{Tsqr}
\end{eqnarray}%
Using $T^{2}=\mathds{1}$ and that the two ends of the segment are
independent, it follows that
\begin{equation}
U^{A}(U^{A})^{\ast }=U^{B}(U^{B})^{\ast }=\exp (i\kappa )\mathds{1},\ \kappa
=0,\pi  \label{kappa}
\end{equation}%
We can thus distinguish two different phases, corresponding to $\kappa
=0,\pi $. Let us now focus on the consequences for the entanglement
spectrum. Assume that $|\phi _{\gamma}\rangle $ is an eigenstate of $%
\mathcal{H}_{S}$ with eigenvalue $E _{\gamma}$, then $U^{A}|\phi
_{\gamma}\rangle $, $U^{B}|\phi _{\gamma}\rangle $, $U^{A}U^{B}|\phi
_{\gamma}\rangle $ are also eigenstates with the same eigenvalue because $%
\mathcal{H}_{S}$ commutes with $U_{A}$ and $U_{B}$. If $\kappa =\pi $, the
unitaries $U^{A},U^{B}$ are anti-symmetric and thus the four states are
mutually orthogonal, resulting in a four fold degeneracy of the entanglement
spectrum. If $\kappa =0$, the entanglement spectrum does not necessarily
have any degeneracies. For example, in the Haldane phase of spin-1 chains,
we find $\kappa =\pi $ and therefore the entire entanglement spectrum of a
segment $S$ is four fold degenerate.\cite{Pollmann-2010}$^,$\footnote{%
Note that Ref. \onlinecite{Pollmann-2010} discusses the entanglement
spectrum at a single cut of an infinite system, and therefore the degeneracy
is two--fold. Here, we are discussing a finite but large segment. There is a
double degeneracy associated with each edge of the segment, so overall, the
entanglement spectrum is \emph{four--}fold degenerate.}

This method may be generalized to give a classification of phases with any
given set of symmetries. For each relationship between the physical
symmetries (e.g., $T^2=\mathds{1}$ in the case just described), there is a
corresponding relationship between the factored symmetries of the
entanglement spectrum\cite{us2048}, in which certain phases (e.g., $\kappa$)
can appear. Certain combinations of these phases are ``gauge invariant"
(independent of how the phases of the factored symmetries are chosen). These
combinations distinguish between topological phases. In fermionic models, an
additional possibility is that symmetry operators at opposite ends may
either commute or anticommute, as described in the previous section.

\paragraph*{Fermionic models.}

We now consider a Hamiltonian which has both fermion parity conservation
with $Q^{2}=\mathds{1}$ and time reversal symmetry with $T^{2}=\mathds{1}$.
In the presence of both symmetries, we show that each of the two phases
defined in the previous section ($\mu =0,\pi $) can be subdivided into four
different phases. Furthermore, we discuss the consequences for the
entanglement spectrum in each case. As $T$ simply takes the complex
conjugate (spin degrees of freedom are not considered here), it does not
change the total fermion number and thus $\left[ T,Q\right] =0$. We will now
classify the phases by examining how the properties of the factored versions
of $Q$ and $T$ depart from the relations of the full transformations, $T^{2}=%
\mathds{1} $, $[T,Q]=0$.

We first consider the case $\mu =0$, \emph{i.e.}, $\left[
Q^{A},Q^{B}\right] =0$. Then, both $Q^{A}$ and $Q^{B}$ are bosonic
operators, and can be
diagonalized simultaneously. Then, we distinguish two cases: $%
Q^{A}T=e^{i\phi }TQ^{A}$ with $\phi =0,\pi $, and similarly for
$Q^{B}$.
Note that $\phi $ has to be the same for $Q^{A}$ and $Q^{B}$, since $%
Q=Q^{A}Q^{B}$ satisfies $\left[ T,Q\right] =0$. If $\phi =\pi $,
time reversal changes the parity of the fermion number in either
end. (A similar situation occurs at the vortex cores of
time-reversal invariant topological
superconductors\cite{Qi-topsc}.) We now examine the two cases
$\phi =0,\pi $ separately.

$(\mu =0\emph{,}$\emph{\ }$\phi =0\emph{,}$\emph{\ }$\kappa =0$ or $\pi )$--
The case $\phi =0$ is analogous to the bosonic case considered above, with
two phases, one corresponding to $\kappa =\pi $, characterized by a
four-fold degenerate entanglement spectrum of a segment, and one to $\kappa
=0$, in which there is no necessary degeneracy in the entanglement spectrum.

$(\mu =0\emph{,}$\emph{\ }$\phi =\pi ,$\emph{\ }$\kappa =0$ or $\pi )$-- If $%
\phi =\pi $, $U^{A}$ and $U^{B}$ (defined through $T=U^{A}U^{B}$; see App. %
\ref{app:qs}) are both fermionic operators, since they change the fermion
parity. We know that $T^{2}=U^{A}U^{B}\left( U^{A}U^{B}\right) ^{\ast
}=-U^{A}U^{A\ast }U^{B}U^{B\ast }=\mathds{1}$. This can only be satisfied if
$U^{A}(U^{A})^{\ast }=\exp (i\kappa )\mathds{1}$ and $U^{B}(U^{B})^{\ast
}=-\exp (i\kappa )\mathds{1}$ with $\kappa =0,\pi $. Note that $\left\{ Q^{A}%
\text{,}U^{A}\right\} =0$ and $\left\{ Q^{B}\text{,}U^{B}\right\}
=0$, where both $U^{A}$ and $U^{B}$ commute with
$\mathcal{H}_{S}$. ($\{\cdot,\cdot\}$ denotes an
anti--commutator.) Therefore, each entanglement level is
four--fold degenerate, where the degeneracy corresponds to states
with all possible combinations of $Q^A=\pm 1$ and $Q^B=\pm 1$.

Next, we consider the case $\mu =\pi $. In this case, $\left\{
Q^{A},Q^{B}\right\} =0$, so these two operators cannot be diagonalized
simultaneously. Rather, every entanglement eigenstate can be labelled by the
eigenvalue $q=\pm 1$ of the parity operator $Q=-iQ^{A}Q^{B}$, where $|\alpha
\beta ,q=\pm 1\rangle $ are degenerate. Since $\left[ T,Q\right] =0$, we
must have either $\left[ T,Q^{A}\right] =0$ and $\{T,Q^{B}\}=0$, or vice
versa. Therefore, we define a parameter $\phi =0,\pi $ such that $%
TQ^{A}=e^{i\phi }Q^{A}T$ and $TQ^{B}=e^{i(\phi +\pi )}Q^{B}T$.

In this case ($\mu=\pi$), phases with $\phi=0$ and $\pi$ behave very
similarly. To see this, we just note that if $\phi=\pi$ then $QT$ commutes
with $Q^A$. Therefore, we will define a modification of time reversal that
commutes with $Q^A$, $T^{\prime}:=QT$ if $\phi=\pi$ and $T^{\prime}:=T$ if $%
\phi=0$. Let the factors of $T^{\prime}$ be $T^{\prime}=U^{A^{\prime}}U^{B^{%
\prime}}$. One can check that $U^{A^{\prime}},U^{B^{\prime}}$ are bosonic.
The entanglement spectrum can be divided into two sectors with a fixed value
of $q$. The operator $U^{A^{\prime}}$, being bosonic, depends only on $%
\alpha,\beta$ and hence acts the same way on both sectors. Define $\kappa$
by $e^{i\kappa}\mathds{1}=U^{A^{\prime}}U^{A^{\prime}*}=U^{B^{\prime}}U^{B^{%
\prime}*}$. The possible values for $\kappa$ are $0$ and $\pi$:

$(\mu =\pi \emph{,}$\emph{\ }$\phi =0$ or $\pi ,$\emph{\ }$\kappa =0$\emph{)}%
-- If $\kappa =0$, each entanglement eigenstate in each $q$ sector can be
singly degenerate. Therefore, counting the $q=\pm 1$ degeneracy, each
entanglement eigenstate has a minimal degeneracy of $2$.

$(\mu =\pi \emph{,}$\emph{\ }$\phi =0$ or $\pi ,$\emph{\ }$\kappa
=\pi )$-- If $\kappa =\pi $, the spectrum in each of the $\pm q$ sectors is
fourfold degenerate (for the same reasons as in the bosonic case above with $%
\kappa =\pi $, \emph{i.e.}, there is a Kramer's doublet at each edge).
Taking the $q=\pm 1$ degeneracy into account, every entanglement eigenstate
is at least eight fold degenerate.

There are therefore eight different phases classified
by triplets $(\mu ,\phi ,\kappa )$%
, where each entry is $0$ or $\pi $. $\mu =\pi $ if there are Majorana modes
at the ends of the segment, $\phi =\pi $ if time reversal and $Q^{A}$
anticommute, and $\kappa =\pi $ if $U^{A}(U^{A})^{\ast }=-\mathds{1}$
(leading to Kramers' doublets at the edges of the
segment). Since, as long as the time-reversal and fermion parity symmetries
are preserved, $(\mu ,\phi ,\kappa )$ can only take the values $0$ or $\pi $%
, they cannot change smoothly; the only way for them to change is through a
non-analytic change of the ground state wave function, \emph{i.e.} a quantum
phase transition.

The eight different phases and their corresponding minimal degeneracies are
summarized in Table \ref{table} in Section~\ref{sec:sum}. The degeneracies
illustrate a distinction between interacting and noninteracting systems. As
we will show in Sec.~\ref{sec:Addition}, the eight distinct phases can be
realized by taking $M$ copies of a single chain in the large $t$ phase,
where $M=1,\dots ,8$. Without interactions, the degeneracy of the Schmidt
spectrum would be equal to $2^{M}$. Interactions can partly lift this
degeneracy but cannot connect the eight phases defined by $\left( \mu ,\phi
,\kappa \right) $ adiabatically.

\section{Addition of Phases}

\label{sec:Addition}The eight phases we have just obtained obey a group
structure, which is defined by the rules of \textquotedblleft
adding\textquotedblright\ them together. This group turns out to be $\mathbb{%
Z}_{8}$. The addition rules reveal interesting distinctions between bosons
and fermions. We will work out the addition table in some detail.

Two systems can be added together by placing them side by side: hence if one
system is in phase $P_{1}$ and another is in $P_{2}$, then the combined
\textquotedblleft ladder\textquotedblright\ system is in phase $P_{1}+P_{2}$%
. (The combined system then remains in $P_{1}+P_{2}$ even when the two
constituent systens are coupled, as long as the coupling Hamiltonian is
symmetric under time-reversal and fermion parity, and the bulk gap does not
collapse.) This rule creates a finite group for a given set of symmetries.
In particular, every element in this group has an inverse and the trivial
phase is the identity element.

The distinction between fermionic and bosonic systems is related to the
inverse operation. If the system consists only of bosons, then the inverse
element of any phase $P$ is its complex conjugate (\emph{i.e.}, its time
reversal):
\begin{equation}
P+P^{\ast }=\mbox{trivial phase}.  \label{rebirth}
\end{equation}%
For an example, consider a spin one Heisenberg chain. A single chain cannot
be adiabatically connected to the trivial phase because its ends transform
as spin 1/2 degrees of freedom. However, as shown in Ref.~[%
\onlinecite{Todo-2001}], two coupled chains can be connected continuously to
the rung singlet phase (i.e., a product state of spin zeros on the rungs).
The two chains are no longer distinguished from the trivial phase by their
ends because the two half-integer spins couple to form integer spin states.

In general, phases of bosonic chains are distinguished by the projective
representation of the symmetry groups acting on the entanglement eigenstates
(see Ref.~\onlinecite{Pollmann-2010}). Each element in the symmetry group $%
\sum $ is represented in the entanglement eigenbasis as a left-hand unitary
matrix $U^{A}\left( \Sigma \right) $ acting on the left index of the state,
and a right-hand matrix $U^{B}\left( \Sigma \right) $ acting on the right
index. Then the combined operation of two elements $\Sigma _{1}$ and $\Sigma
_{2}$ is represented by $U^{A}\left( \Sigma _{1}\Sigma _{2}\right) =e^{i\rho
_{A}\left( \Sigma _{1},\Sigma _{2}\right) }U^{A}\left( \Sigma _{1}\right)
U^{A}\left( \Sigma _{2}\right) $, and similarly for $U^{B}$. 
To see that $P+P^{\ast }$ is trivial, consider the eigenstates of the
entanglement Hamiltonian for a segment in the combined system
\begin{equation}
|\alpha \beta \rangle _{\text{coupled}}=|\alpha _{1}\beta _{1}\rangle
_{P}|\alpha _{2}\beta _{2}\rangle _{P^{\ast }}.
\end{equation}%
The left-hand matrix $U^{A}$ representing a symmetry $\Sigma $ for the
coupled system is
\begin{equation}
U_{\alpha _{1}^{\prime }\alpha _{2}^{\prime };\alpha _{1}\alpha _{2}}^{A,%
\text{coupled}}(\Sigma )=U_{\alpha _{1}^{\prime }\alpha _{1}}^{A}(\Sigma
)U_{\alpha _{2}^{\prime }\alpha _{2}}^{A\ast }(\Sigma )  \label{roller}
\end{equation}%
where the second factor is complex-conjugated because the second chain is
time-reversed. Hence, the phase factors cancel, $U^{A,\text{coupled}}(\Sigma
_{1})U^{A,\text{coupled}}(\Sigma _{2})=U^{A,\text{coupled}}(\Sigma _{1}\Sigma
_{2})$, and the resulting system is in a trivial phase.

Now we can try to build more complicated phases out of simpler ones by
placing them side-by-side. For bosonic systems,
this procedure does not generate new phases in the presence
of time reversal symmetry.
Time reversal symmetry and Eq. (\ref{rebirth})
imply that $P+P=0$. Hence
starting from one phase, it is not possible to get more than two phases (the
original phase and the trivial phase). There may be additional phases that
would have to be built up from independent starting points. The group is
always a product of $\mathbb{Z}_2$'s, in other words.

However, for fermionic spin chains, $P$ and $P^{\ast }$ are not necessarily
inverses. The $p$-wave superconducting state $P_{1}$ described by Eq. (\ref%
{singlechain}) with $t>\mu $ which is an order eight phase, as discovered by
Fidkowski and Kitaev\cite{Fidkowski-2010}, is an illustration. Eq. (\ref%
{roller}) breaks down because operators on the two chains can anticommute
with each other. In fact, starting from a single Majorana chain, we can
generate all possible combinations of $\mu $, $\phi $, and $\kappa $. We now
demonstrate this idea for a number of examples:

\begin{enumerate}
\item Consider the Majorana chain, the ground state of Eq. ~(\ref%
{singlechain}) with $t>u$ which is in the $(\mu ,\phi ,\kappa )=(\pi ,0,0)$
phase. When two copies are combined together, the resulting phase has $(\mu
,\phi ,\kappa )_{\text{coupled}}=(0,\pi ,0)$, i.e., the ends are not
Majorana fermions any more, but time reversal changes the fermion parity of
the ends. The fermion parity for the segment of the combined chain is given
by $%
Q=Q_{1}Q_{2}=(iQ_{1}^{A}Q_{1}^{B})(iQ_{2}^{A}Q_{2}^{B})=(Q_{1}^{A}Q_{2}^{A})(Q_{1}^{B}Q_{2}^{B})
$, where the $Q_{n}^{A},Q_{n}^{B}$ are fermionic parity operators of a chain
$n$ with Majorana ends (see Sec. \ref{parrrr}). One can measure the parities
of the ends separately because $Q^{A}=-iQ_{1}^{A}Q_{2}^{A}$ and a similar
operator on $B$ are bosonic operators, so $\mu _{\text{coupled}}=0$. On the
other hand $\phi _{\text{coupled}}=\pi $ because $T$ anti-commutes with $%
Q^{A}$ on account of the factor of $i$. Furthermore, one finds that $\kappa
_{\text{coupled}}=\pi$. (See App.~\ref{app:qs}.)

\item Consider two chains with $(\mu ,\phi ,\kappa )=(0,\pi ,\pi)$. $%
Q_{1}^{A}$, $Q_{2}^{A}$ are bosonic, therefore $Q_{\text{coupled}%
}^{A}=Q_{1}^{A}Q_{2}^{A}$ is bosonic as well, and $\mu _{\text{coupled}}=0$.
Time reversal acting on the left edge is represented as $U_{\text{coupled}%
}^{A}=U_{1}^{A}U_{2}^{A}$. It anti-commutes with $Q_{1}^{A}$, $Q_{2}^{A}$,
but commutes with their product, therefore $\phi _{\text{coupled}}=0$. Since
both $U_{1}^{A}$, $U_{2}^{A}$ change the fermion parity, they both have to
be \emph{fermionic}. Therefore $\left\{ U_{1}^{A},U_{2}^{A}\right\} =0$ and
we get that%
\begin{eqnarray}
U_{\text{coupled}}^{A}U_{\text{coupled}}^{A\ast } &=&\left(
U_{1}^{A}U_{2}^{A}\right) \left( U_{1}^{A}U_{2}^{A}\right) ^{\ast }
\nonumber \\
&=&-U_{1}^{A}U_{1}^{A\ast }U_{2}^{A}U_{2}^{A\ast }  \nonumber \\
&=&-\mathds{1}\text{.}
\end{eqnarray}
Hence $\kappa _{\text{coupled}}=\pi $, and the resulting phase is labelled
by $\left( 0,0,\pi \right) $.

\item Combining two chains with $(0,0,\pi )$ finally gives the trivial
phase, because all the symmetries are represented by bosonic operators,
therefore $\kappa $ simply doubles to give $0\ \mathrm{mod}\left( 2\pi
\right) $. This conforms with the fact that the Majorana chain is an order
eight element of the group.
\end{enumerate}

Working out the addition rule in general gives the table of phases which are
summarized in Table \ref{table}. A concise way to describe the general
addition rule is to define $\lambda \equiv\kappa +\phi (\mathrm{mod\ }2\pi)$%
. Then we represent a state by a 3-digit binary number $\left( \frac{\lambda
}{\pi },\frac{\phi }{\pi },\frac{\mu }{\pi }\right) $. These numbers add
modulo 8 when the phases are combined.

\section{Translational Invariance and $\protect\theta $}

\label{sec:trans} If, in addition to fermion parity conservation,
translational invariance is also present, the number of distinct phases is
doubled. Below, we derive the associated invariant, $\theta $, which can
take the values $0$ or $\pi $, independent of the invariants $\left( \mu
,\phi ,\kappa \right) $ described above. The degeneracy of the entanglement
spectrum, however, is not modified in either the $\theta =0$ or $\pi $
phases, and is given by Table \ref{table}.

Let us consider a fermionic chain with translational invariance. According
to the Sec. \ref{parrrr}, the fermion parity of a segment $S=\left[ 1,L%
\right] $ extending from $j=1$ to $j=L$ (where $L$ is much larger than the
correlation length $\xi $) can be written as
\begin{equation}
Q\left( 1,L\right) =e^{-i\frac{\mu }{2} }f\left( L\right) Q^{A}\left(
1\right) Q^{B}\left( L\right) \text{.}  \label{Q1L}
\end{equation}%
Here, we have kept track explicitly of the position of the operators $Q^{A}$
and $Q^{B}$, and of an overall constant sign $f\left( L\right) $ (which was
absorbed into the definition of $Q^{A}$ and $Q^{B}$ before). Translational
invariance removes the necessity of choosing the sign of $Q^B$ (and $Q^A$)
separately
for each segment.
This symmetry also allows us to write the parity operator of the
segment $\left[ 1,L\right] $ in terms of those of the two segments $S_{1}=%
\left[ 1,L^{\prime }\right] $ and $S_{2}=\left[ L^{\prime }+1,L\right] $
(where $L^{\prime},L-L^{\prime}\gg \xi $) as
\begin{eqnarray}
Q\left( 1,L\right) &=&Q\left( 1,L^{\prime }\right) Q\left( L^{\prime
}+1,L\right)  \nonumber \\
&=&e^{-i\frac{\mu }{2} }f\left( L^{\prime }\right) f\left( L-L^{\prime
}\right)  \label{Q1L_div} \\
&&\times Q^{A}\left( 1\right) \left[ e^{-i\frac{\mu }{2} }Q^{B}\left(
L^{\prime }\right) Q^{A}\left( L^{\prime }+1\right) \right] Q^{B}\left(
L\right) \text{.}  \nonumber
\end{eqnarray}%
 Equating Eq. \ref{Q1L}
and \ref{Q1L_div} gives that, within the low--entanglement subspace, we must
have
\begin{equation}
f\left( L^{\prime }\right) f\left( L-L^{\prime }\right) \left[ e^{-i\frac{\mu
}{2} }Q^{B}\left( L^{\prime }\right) Q^{A}\left( L^{\prime }+1\right) %
\right] =f\left( L\right) \text{.}  \label{equality}
\end{equation}%
Eq. \ref{equality} can hold for every state in the low-entanglement subspace
only if these states are all eigenstates of $e^{-i\frac{\mu }{2}
}Q^{B}\left( L^{\prime }\right) Q^{A}\left( L^{\prime }+1\right) $. $Q^{A}$
and $Q^{B}$ can be defined in such a way that the corresponding eigenvalue
is $1$. Then, we get that $f\left( L^{\prime }\right) f\left( L-L^{\prime
}\right) =f\left( L\right) $, which is solved by
\begin{equation}
f\left( L\right) =e^{i\theta L}\text{.}
\end{equation}
From the requirement that $\left[ Q\left( 1,L\right) \right] ^{2}=\mathds{1}$%
, we get that $\theta $ can only take the values $0$ or $\pi $. Thus, each
of the eight phases found in the previous section is further split into two
distinct phases, corresponding to the two allowed values of $\theta $. For
example, for the Majorana chain model (Eq. \ref{singlechain}), $t=0$, $u=+1$
and $t=0$, $u=-1$ describe distinct phases, although both have $\mu =0$. The
ground state has all sites occupied or unoccupied, corresponding to $\theta
=\pi $ or $\theta =0$, respectively.

Note that both $\theta$ and $\mu$ have a concrete consequence not only for
the entanglement spectrum but also for the parity of the ground state in
periodic chains.
If the length of the chain is much larger than the correlation length, the
parity depends only on $\mu$ and $\theta$ and the chain length, $%
(-1)^{(\mu+\theta L)/\pi}$. Thus, a phase with $\mu=\pi$ has an odd number
of fermions on a chain of an even length.\cite{Kitaev-2001} The phase $%
\theta $ determines whether the parity of the ground state alternates as a
function of $L$ or not. This is shown in appendix \ref{app:parity}.

\section{Summary and outlook\label{sec:sum}}

We have described a systematic procedure for classifying the phases of 1D
interacting fermions. We focussed on spinless fermions with time-reversal
symmetry and particle number parity conservation. In the non-interacting
case, these models are classified by an integer number, i.e., by $\mathbb{Z}$%
.\cite{Schnyder-2008,Kitaev-2009} We used concepts of entanglement
to classify the phases in the presence of interactions. We derive
an effective description of the dominant entanglement states which
then allows us to recognize ``topological'' features based on
projective representations of the symmetries. We found, in
agreement with the results of Fidkowski \emph{et. al.} (Ref.
[\onlinecite{Fidkowski-2010}]), that in the presence of
interactions there are only eight distinct phases. Each of these
eight phases is characterized by a unique set of bulk invariants
$(\mu ,\phi ,\kappa )$, which can take the values $0$ or $\pi $.
These invariants are related to the transformation laws of the
entanglement eigenstates under symmetry operations, and the phases
have a characteristic degeneracy of entanglement levels.
Furthermore, the phases obey a $\mathbb{Z}_{8}$ group structure
and each of the eight phases can be generated by adding single
chains together. All possible phases and the addition rules are
summarized in Table~\ref{table}. If translational symmetry is also
present, the number of distinct phases increases to 16.

The symmetries we have focused on describe only one of the 10
Altland-Zirnbauer classes\cite{Altland1997} of topological
insulators. The framework described here can also be used to show
how the phases in these other classes are modified by
interactions. To analyze each of the classes of topological
insulators (and in both interpretations when particle-hole
symmetry is present), one only has to determine the appropriate
algebra of symmetries and then determine the possible projective representations
of this algebra.

 Interactions cause the meaning of the Altland-Zirnbauer
classes to bifurcate, however. At the mean-field level, a
superconductor has an emergent particle-hole symmetry in its band
structure. Thus, the classes which have such a symmetry can be
interpreted as describing either superconductors or systems that
have a true particle-hole symmetry (such as the Hubbard model for
fermions with spin on a
bipartite lattice at half-filling). When interactions are
included, these two interpretations are distinct.  Thus, the class
BDI, for example, has particle-hole and time reversal symmetry.
This can be interpreted as describing superconductors.  This means
that one symmetry, particle conservation, breaks down, and only
fermion parity $Q$ is left. The only two symmetries are $T$ and
$Q$, giving the problem treated here. BDI has an alternative
interpretation, according to which it describes
systems with a true particle-hole symmetry $C$ that
reverses the sign of $\langle n_i\rangle-\bar{n}$. (Here,
$n_i$ is the occupation number of a site and $\bar{n}$ is
the mean occupation number.) In this case, particle
number $N$ is conserved, and besides this there are two other
symmetries $T$ and $C$. These satisfy the algebra
$T^2=C^2=\mathds{1}$, $CN+NC=2\bar{n} L$ where $L$ is the length
of the system.
(Every other pair of these symmetries commute.) The set of phases
is different for the two interpretations; particle number
conservation rules out Majorana Fermions. A complete
classification of systems in all Altland-Zirnbauer classes in one dimension,
following either of the two interpretations of particle-hole
symmetry mentioned above, would be an interesting project for future work.

A generalization of these results to higher dimensional systems is an
interesting (and challenging) open problem. In some of the symmetry classes
of topological insulators and superconductors, strong arguments have been
given that the non--interacting classification does not change when
interactions are included. This is particularly clear when the topological
invariant is related to a quantized physical response, \emph{e.g.}, in the
integer quantum Hall effect\cite{NiuThouless, Avron1985} and in 3D
time--reversal invariant topological insulators\cite{Qi-2008}. However, for
other classes, the situation is less clear. For example, the
non--interacting classification of 3D chiral superconductors is $\mathbb{Z}$,%
\cite{Schnyder-2008,Kitaev-2009} similar to the one--dimensional case
considered here. It would be interesting to consider the effect of
interactions on the phase diagrams of such systems.
\begin{table}[tbp]
\begin{tabular}{c|c|c}
Number of chains & $(\mu,\phi,\kappa)$ & Degeneracy of segment \\ \hline
1 & $(\pi,0,0)$ & 2 \\
2 & $(0,\pi,\pi)$ & 4 \\
3 & $(\pi,\pi,\pi)$ & 8 \\
4 & $(0,0,\pi)$ & 4 \\
5 & $(\pi,0,\pi)$ & 8 \\
6 & $(0,\pi,0)$ & 4 \\
7 & $(\pi,\pi,0)$ & 2 \\
8 & $(0,0,0)$ & 1%
\end{tabular}%
\caption{Degeneracies and addition table. All possible phases of fermions
are realized by simply taking copies of some number of Majorana chains (see
next section); the first column is the number of chains. The next column
gives the parameters classifying a given state. The third column gives the
degeneracy of the Schmidt spectrum, counting \emph{both} ends.}
\label{table}
\end{table}

\emph{Note} As we were writing this article, we learned that a similar
classification is being worked out by Fidkowski and Kitaev.\cite%
{Fidkowski-2010b} Our results are consistent with theirs.

\section*{Acknowledgment}

We thank Lukasz Fidkowski and Shinsei Ryu for useful discussions. E.~B. was
supported by the NSF under grants DMR-0705472 and DMR-0757145. F.~P. and
A.~M.~T. acknowledge support from ARO grant W911NF-07-1-0576.

\appendix

\section{Low Energy Operators and the Ends of the Chain}

\label{app:operators} An intuitive argument, given above, suggests that
low-energy operators acting on the entanglement eigenstates may be
represented approximately by operators located near the ends of the chain:
in each Schmidt state $|\alpha\beta\rangle_S$ the expectation values of the
spin and other operators have some particular spatial dependence near the
ends of the chain depending on $\alpha$ and $\beta$, but this decays
exponentially to the ground state away from the ends. Therefore it should be
possible to transform between these states by using operators defined on
just the ends. A special case is the effective representations of symmetries
in terms of operators at the ends of the segments, which we used to define
the topological phases.

To see that these effective operators exist, one can use a matrix product
state representation\cite{Fannes-1992} of the wave function. (We focus here just on bosonic
systems. For systems including fermions a similar argument can be developed
using bosonization but the discussion of this general result gets
complicated by the classification of the phases by $\mu$. )

A basis for the low-energy operators can be constructed as follows. For each
fixed choice of $\alpha _{1}$ and $\alpha _{2}$ let $\mathcal{O}^{A}(\alpha
_{2},\alpha _{1})$ be the operator that transforms $\alpha _{1}$ into $\alpha
_{2}$.

We will now give an approximate representation for $\mathcal{O}^{A}(\alpha
_{1},\alpha _{2})$ that gives the correct transformation of \emph{low energy
states} of $\mathcal{H}_S$. Let the matrices $\Gamma _{m},\Lambda $ define the bulk state ($m$
varies over a basis for the physical Hilbert space). The ground state
wavefunction of a ring of length $N$ is given by

\begin{equation}
|\psi \rangle =\sum_{\{m_{i}\}}\mathrm{tr}(\Gamma _{m_{1}}\Lambda \Gamma
_{m_{2}}\dots \Lambda \Gamma _{m_{N}} \Lambda)|m_{1}m_{2}\dots m_{N}\rangle
\label{GS}
\end{equation}

$\Gamma _{m},\Lambda $ can be brought into a canonical form, satisfying $%
\sum_{m}\Gamma _{m}\Lambda \Gamma _{m}^{\dagger }=\sum_{m}\Gamma
_{m}^{\dagger }\Lambda \Gamma _{m}=\mathds{1}$, where $\Lambda $ is a
diagonal matrix with non--negative entries.\cite{Vidal-2003a, Orus-2008} For
a generic wave function, $\Gamma _{m}$, $\Lambda $ are infinite-dimensional.

Let us define the states $|\alpha \beta \rangle _{1L}$ of a segment of the
chain stretching from $1$ to $L$, where $L<N$, as%
\begin{equation}
|\alpha \beta \rangle _{1,L}=\sum_{\{m_{i}\}}(\Gamma _{m_{1}}\Lambda \Gamma
_{m_{2}}\dots \Lambda \Gamma _{m_{L}})_{\alpha \beta }|m_{1}m_{2}\dots
m_{L}\rangle  \label{eq:traincar}
\end{equation}

When $L$ is large these states are nearly orthonormal, that is $|\langle
\alpha ^{\prime }\beta ^{\prime }|\alpha \beta \rangle _{1,L}-\delta
_{\alpha ^{\prime }\alpha }\delta _{{\beta ^{\prime }\beta }}|\sim C e^{-\frac{%
L}{\xi }}$, where $\xi $ is the length scale for the decay
and $C$ is a constant dependending on the indices. In this limit, $%
|\alpha \beta \rangle _{1L}$ are the entanglement eigenstates of the
segment. On the other hand, if the length of the chain is fixed and $\alpha
,\beta ,\alpha ^{\prime },\beta ^{\prime }$ increase, the orthonormality
must eventually break down for high enough $\alpha ,\beta ,\alpha ^{\prime
},\beta ^{\prime }$ (since the Hilbert space of a segment of length $L$ is
finite). Indeed, $C$ grows as a function of $\alpha,\alpha',\beta,\beta'$.

The ground state wavefunction (\ref{GS}) can now be written as%
\begin{equation}
|\psi \rangle =\sum_{\alpha ,\beta }\lambda _{\alpha }\lambda _{\beta
}|\alpha \beta \rangle _{1,L}|\alpha \beta \rangle _{L+1,N}\text{.}
\end{equation}%
This gives the Schmidt decomposition into states of the environment $|\alpha
\beta \rangle _{L+1,N}$ and states of the chain $|\alpha \beta \rangle
_{1,L} $, with a Schmidt eigenvalue $\lambda _{\alpha }\lambda _{\beta }$
[or, equivalently, an entanglement energy
$E=E_\alpha+E_\beta=-2(\ln \lambda _{\alpha }+\ln
\lambda _{\beta })$]. The Schmidt eigenstates become orthogonal to each
other in the limit $N\rightarrow \infty $ and $L\rightarrow \infty $.

We can now give an \textquotedblleft effective\textquotedblright\ expression
for $\mathcal{O}^{A}$ in terms of local operators acting on sites $1,\dots
,\ell $ (near the left edge of the segment $\left[ 1,L\right] $), valid for
a low entanglement-energy subspace with
$E_\alpha<E_{\text{cut}} $. $E_{\text{cut}} $ is a
cutoff which depends on $\ell $. The accuracy of our effective expression
improves as $\ell $ becomes larger (provided that $N\gg \ell $). Define
\begin{equation}
\mathcal{O}_{\alpha _{2}\alpha _{1}}^{A,\mathrm{eff}}=\sum_{\gamma =1}^{\chi
}|\alpha _{2}\gamma \rangle _{1,\ell }\langle \alpha _{1}\gamma |_{1,\ell }
\label{eq:answer}
\end{equation}%
where $\chi $ is a cutoff of the entanglement spectrum that satisfies $%
E_{\gamma =1,\dots ,\chi }<E_{\text{cut}} $. $\mathcal{O}_{\alpha _{2}\alpha
_{1}}^{A,\mathrm{eff}}$ acts only on the $\ell $ first sites of the segment $%
\left[ 1,L\right] $.

We now apply $\mathcal{O}_{\alpha _{2}\alpha _{1}}^{A,\mathrm{eff}}$ to the
state Eq. ($\ref{eq:traincar}$), with $\alpha \leq \chi $. This state can be
expanded $\sum_{\gamma ^{\prime }}|\alpha \gamma ^{\prime }\rangle _{1\ell
}\lambda _{\gamma }|\gamma ^{\prime }\beta \rangle _{\ell +1,N}$. Using the
approximate orthornormality of the states on the segment from $1$ to $\ell $%
, we find that the operator in fact transforms $\alpha _{1}$ into $\alpha
_{2}$. Intuitively, the sum
over $\gamma $, the state of the internal end of the $\ell -$site segment
ensures that this operator keeps the right type of entanglement between the
left and right side of the \textquotedblleft cut" at $\ell $. The error of (\ref{eq:answer}) scales
as $F(\chi )e^{-\frac{\ell }{\xi }}$, where $F\left( \chi \right) $ is a
function of $\chi $.  ($F(\chi)$ grows with $\chi$, hence,
to deal with a larger range of ``energies", a larger value of $\ell$ must
be used. This is because, as the ``energy" of a state increases,
it penetrates further into the bulk.)

Note that $\mathcal{O}_{\alpha _{2}\alpha _{1}}^{A,\mathrm{eff}}$ does not
transform high--entanglement energy states correctly, that is states such as
$|\alpha \beta \rangle _{1N}$ with $\alpha >\chi $. This is because $|\alpha
_{1}\gamma \rangle $ and $|\alpha \gamma ^{\prime }\rangle $ are
orthonormal only if they are both low energy states. This limitation is
unavoidable: it is not possible to find a perfect representation for an
operator, such as $Q$, in terms of just the $\ell $ sites near each end. One
can add an extra particle somewhere outside of the reach of these sites,
changing the value of $Q$ but not of an observable on the ends of the chain.
The \emph{physical} energy of this state may not be much greater than the
gap. However, not being able to describe states like this is not a problem
when one is studying the ground state of the system: its \emph{entanglement}
energy is large, which means that it contributes negligibly to the value of
any observable in the ground state.

\section{Factoring antiunitary operators\label{app:qs}}

In the analysis of time--reversal symmetry, we defined a parameter $\kappa $
by factoring $T$ into two operators $U^{A}$ and $U^{B}$, acting near the two
opposite edges of the segment. To determine how chains add to one another,
it is necessary to know the commutation and anticommutation properties of
these operators. We ignored a small detail, however: since $T$ is
antiunitary, it cannot be factored either as the product of two unitary or
two antiunitary operators. One solution is just to explicitly write how $T$
transforms the basis states as we did in Eqs. (\ref{timefactorization}) and (%
\ref{Tsqr}). This becomes cumbersome after a while however, and later in the
text we have treated $U^{A}$ and $U^{B}$ as unitary operators in Hilbert
space, without keeping explicit track of their indices. Here we will explain
the meaning of this.

We will first discuss the bosonic case. Eq. (\ref{timefactorization}) gives
the action of $T$ only on basis states. Taking a superposition gives a
factorization of $T$ that is correct for any state:
\begin{equation}
T=U_{A}U_{B}K\text{,}  \label{refactory}
\end{equation}%
where $U_{A}$ and $U_{B}$ are \emph{unitary} operators at the two ends and $%
K $ is defined by
\begin{equation}
K\sum_{\alpha \beta }a_{\alpha \beta }|\alpha \beta \rangle =\sum_{\alpha
\beta }a_{\alpha \beta }^{\ast }|\alpha \beta \rangle \text{,}  \label{K}
\end{equation}%
where $a_{\alpha \beta }$ are arbitrary coefficients. We can now define $%
\kappa $ by $(U^{A}K)^{2}=e^{i\kappa }\mathds{1}$. Thus, $U^{A}K$ is an
antiunitary symmetry squaring to $-\mathds{1}$ in the nontrivial phase as in
Kramers' theorem. Note that this equation is equivalent to the definition
given above, Eq. (\ref{kappa}). This is because $KU^{A}K^{-1}=U^{A\ast }$
when the matrices are represented in the basis $|\alpha \beta \rangle $.
(Note that the complex conjugate of a matrix depends on the basis being
used, unlike the adjoint.)

We can argue physically that operators $U^{A}$ and $U^{B}$ satisfying Eq. (%
\ref{refactory}) can always be found. Consider the ratio $TK^{-1}$ between $%
T $, which is represented by complex conjugation in terms of the microscopic
degrees of freedom, and $K$ which describes complex conjugation in the
entanglement eigenstate basis. This operator is unitary. Furthermore it acts
independently on the two ends: one may check that $K\mathcal{O}^{A,B}K$ is
an operator acting on end $A$ or $B$ respectively, by expressing it in the
basis of entanglement eigenstates.

The operator $U^{A}K$ used to define $\kappa $ is non--local. It does not
commute with operators at end $B$, because it takes complex conjugates of
them. However, we can still argue that $\kappa $ is well-defined: Square
Eq.~(\ref{refactory}): $\mathds{1}=U^{A}U^{B}KU^{A}U^{B}K$. Since $K^{2}=%
\mathds{1}$, we can write this also as $\mathds{1}%
=U^{A}U^{B}(KU^{A}K)(KU^{B}K)$. $KU^{A}K$ is an operator which acts on end $%
A $, therefore it commutes with $U^{B}$. Hence $\mathds{1}%
=[U^{A}(KU^{A}K)][U^{B}(KU^{B}K)]$. Since the two factors are local, each
must be a pure phase, hence $(U^{A}K)^{2}=e^{i\kappa }\mathds{1}$.

Now the operators $U^A$ and $U^B$ are not uniquely defined because \emph{%
complex conjugation}, $K$, \emph{is basis dependent}. Changing the basis of
eigenstates in which Eq. (\ref{K}) is imposed (e.g., multiplying the
entanglement states by phase factors) changes $K$.
This does not change topological properties like the value of $\kappa$
however: the unitary transformation that changes the basis can be carried
out continuously, starting from the identity. In this process, $\kappa$
cannot change because it can only be $0$ or $\pi$.

For fermionic systems with $\mu =0$, one can decompose $T$ using Eq. (\ref%
{refactory}). When $\mu =\pi $, the situation is more complicated because
the parity eigenvalue $q$ cannot be associated with either one of the edges.
We have to make sure that $K$ still maps operators at each end of the system
to other operators at that end. In particular, $KQ^{A}K$ must be a local
operator at end $A$.

This condition is satisfied if $K$ is defined to be complex conjugation in
the basis $|\alpha \beta q\rangle $ \emph{provided} that $Q^{A}$ and $Q^{B}$
are represented by either purely real or purely imaginary matrices in that
basis. One way to satisfy this requirement is to first choose
a basis for $q=+1$ and then to construct
the states in the sector from them,
 $|\alpha \beta ,q=-1\rangle
=Q^{A}|\alpha \beta ,q=+1\rangle $. Then $Q^{A}$ is represented by $\sigma
^{x}$, acting in the $q=\pm 1$ basis. In this basis, each state is an
eigenstate of $Q=\sigma ^{z}$. Last, $Q^{B}=iQQ^{A}=\sigma ^{y}$. Since $%
Q^{A}$ is real and $Q^{B}$ is imaginary, the two ends are not mixed by
applying $K$. (If the relative phases of the basis states are changed, then
simple complex conjugation would mix $Q^{A}$ and $Q^{B}$ into one another.)
We have taken the convention that $Q^{A}$ is real and $Q^{B}$ is imaginary
above.

Now let us show how to calculate $\kappa$ when two $(\mu,\phi,\kappa)=(%
\pi,0,0) $ chains are combined. The factorization $T=U^AU^BK$
must be carried out in a basis of states of the form $|\alpha\beta\rangle$, according
to our conventions.
One basis for the states on the two chains together is given by
$\left\{|\pm\rangle_{1}|\pm{\rangle}_{2}\right\}$ (where the sign represents
the values of $q_1,q_2$. (We do not explicitly  write the bosonic
 indices $\alpha,\beta$.).
 These states map to themselves under time reversal.  However, $U^A$
 and $U^B$ cannot be the identity because
 we know they must be fermionic; this is the wrong basis for defining
 $K$ by simple complex conjugation.

Let us transform the states to a basis in which there is no
entanglement between the ends; we therefore use states that are eigenvectors
of the local
operators $Q^A=-iQ^A_1Q^A_2$ and $Q^B=iQ^B_1Q^B_2$,
namely $|q_A q_B\rangle$.  (The relative minus sign between $Q^A$ and
$Q^B$ ensures that the total parity is $q_Aq_B=q_1q_2$.)

To construct the basis, first find an eigenfunction of $Q^A$ and $Q^B$ with eigenvalues $+1$,
(choose the phase arbitrarily):
\begin{equation}
|+_A+_B\rangle=\frac{1}{\sqrt{2}}(|+\rangle_1|+\rangle_2-i|-\rangle_1|-%
\rangle_2)
\end{equation}%
 Now generate the other basis states from this by applying $%
Q^A_1$ and $Q^B_1$:
\begin{eqnarray}
|-_A+_B\rangle&=&Q^A_1|+_A+_B\rangle=\frac{1}{\sqrt{2}}(|-\rangle_1|+%
\rangle_2-i|+\rangle_1|-\rangle_2)  \nonumber \\
|+_A-_B\rangle&=&-iQ^B_1|+_A+_B\rangle=\frac{1}{\sqrt{2}}(|-\rangle_1|+%
\rangle_2+i|+\rangle_1|-\rangle_2)  \nonumber \\
|-_A-_B\rangle&=&Q^A_1|+_A-_B\rangle=\frac{1}{\sqrt{2}}(|+\rangle_1|+%
\rangle_2+i|-\rangle_1|-\rangle_2)\nonumber \\
\end{eqnarray}
The phases are just conventions in the first two definitions, and the phase
in the third equation follows from the independence of the ends: $Q^A_1$ has
to act on $q_A$ the same way no matter what the value of $q_B$ is.
Now it is clear that $T$ switches the fermion parity of
each end in this basis,  since changing the sign of $i$ exchanges
the states $|q_Aq_B\rangle$ and $|-q_A,-q_B\rangle$.

Now we can define $K$ to map each of \emph{these} basis states to itself.
Clearly, $T=Q_2$ because $Q_2$ also exchanges
the same pairs of wave functions, or more precisely $T=Q_2K$.
Hence $U^{A}=-iQ^{A}$ and $U^B=Q^B$.  One can check that
$(U^AK)^2=-\mathds{1}$,
so  $\kappa=\pi$.

Note that, in spite of all this trouble, the value of $\kappa $ in a phase
with $\mu =0,\phi =\pi $ does not have any physical significance--the
four-fold degeneracy of the spectrum is already explained by the fact that $%
\phi =\pi $. The reason $T$ changes fermion parity at the ends is that the
two ends can only be disentangled by a change of basis that includes complex
phases.

\section{Parity of the ground state on a periodic chain}

\label{app:parity} The parity of the ground state on a periodic chain is
given by $e^{i(\theta L+\mu)}$. This follows from a fact in Sec. \ref%
{sec:trans}. Consider two subsegments of the chain, one ending at $X$ and
the other starting at $X+1$. The ground state wave function is an
eigenfunction of the following:
\begin{equation}
e^{-i\frac{\mu}{2}}Q^B(X)Q^A(X+1)|\psi\rangle=|\psi\rangle.
\label{eq:homecoming}
\end{equation}
When $\mu=\pi$, this relation describes the correlations between the
Majorana degrees of freedom in adjacent segments of the chain.

We now suppose the periodic chain has length $L$ and break it at two places,
between $L^{\prime}$ and $L^{\prime}+1$ and between $L$ and $1$. The total
fermion parity of the ground state is the product of the parity on the two
segments, $(e^{i(\theta L^{\prime}-\frac{\mu}{2})}Q^A(1)Q^B(L^{%
\prime}))(e^{i(\theta(L-L^{\prime})-\frac{\mu}{2})}Q^A(L^{\prime}+1)Q^B(L))$%
. Rearranging and using Eq. (\ref{eq:homecoming}), the ground state parity
comes out as $e^{i(\theta L+\mu)}$. The extra minus sign when $\mu=\pi$
comes from anticommuting the $Q$ operators.

\bibliographystyle{apsrev1}
\bibliography{bibo}

\end{document}